\documentclass[preprint,review,12pt]{elsarticle}

\newcommand{\beq}{\begin{equation}}
\newcommand{\eeq}{\end{equation}}
\newcommand{\bqa}{\begin{eqnarray}}
\newcommand{\eqa}{\end{eqnarray}}

 \usepackage{graphicx}
\usepackage{epsfig}
\usepackage{amssymb}
 \usepackage{amsthm}

\journal{Nuclear Physics A}

\begin{document}

\begin{frontmatter}

\title{Stability of 1+1 dimensional causal relativistic viscous hydrodynamics}

\author[label1]{J. W. Li}
\author[label1]{Y. G. Ma\corref{cor1}}
\author[label1]{G. L. Ma}

\address[label1]{Shanghai Institute of Applied Physics,
        Chinese Academy of Sciences,\\P. O. Box 800-204, Shanghai 201800,
        China}
\cortext[cor1]{Corresponding author. {\it Email address:}
ygma@sinap.ac.cn(Y. G. Ma)}

\begin{abstract}
The stability of the 1+1 dimensional solution of Israel-Stewart
theory is investigated. Firstly, the evolution of the temperature
and the ratio of the bulk pressure over the equilibrium pressure of
the background is explored. Then the stability with linear
perturbations is studied by using the Lyapunov direct method. It
shows that the shear viscosity may weaken the instability induced by
the large peak of bulk viscosity around the phase transition
temperature $T_c$.

\end{abstract}


\end{frontmatter}

\newpage

\section{Introduction}
\label{}

A hot and dense partonic matter has been created at the Relativistic
Heavy Ion Collider (RHIC) of Brookhaven  National Laboratory
\cite{White-papers}, which is thought to be a kind of perfect liquid
of quark and gluon~\cite{Gyul}. The study on the properties of this
matter is a hot issue. Presently one property which is of very
interest is the exact value of the ratio of viscosity over entropy
density for the matter. To extract the value, one needs to compare
experimental data with relativistic viscous hydrodynamics
simulation~\cite{Roma1,Roma2,Song1,Song2,Chaud}.

Two different kinds of relativistic viscous hydrodynamics have been
developed so far. One is the so-called first order relativistic
viscous hydrodynamics which was first developed by Eckart and varied
by Landau and Lifshitz\cite{Eckart,Landau}, and many calculations
have been done since then\cite{Teany}. But there are two problems in
the approaches: one is that dissipative fluctuation may propagate at
a speed larger than the speed of light and thus leads to a causality
problem; the other is that the solution which may develop
instabilities \cite{Muronga}. These problems have been studied
extensively in Refs. \cite{Hisc1,Koun}. The other is the second
order theories among which the Israel-Stewart theory is popularly
used so far \cite{Isra}. However, before Israel-Stewart theory is
applied to describe heavy ion collisions, one should know whether it
is stable or not. Till now, there have been some stability analysis
on the issue \cite{Hisc2,Deni,Deni2,Baier}. In
\cite{Hisc2,Deni,Deni2,Baier}, the authors used the plane wave
perturbation method to study the stability of the theory around a
hydrostatic state and discussed the regime of validity of
hydrodynamics. Also in \cite{Deni2}, using the Lyapunov direct
method as in \cite{Koun}, the authors analyzed the stability of the
scaling solution with only bulk viscosity and presented the stable
regions of the scaling solution for both homogeneous and
inhomogeneous perturbations. And several groups have compared this
theory with the experimental data and show that this theory can be
used without a problem when the shear viscosity and bulk viscosity
are small \cite{Roma1,Roma2,Song1,Song2,Chaud,Song3}.

It is well known that hydrodynamics fails to describe the HBT
results, i.e, the calculated ratio of HBT radii in outward direction
over that in sideward direction is higher than the experimental
data\cite{Pratt}. Recent attempt to reconcile them is to take bulk
viscosity into account. But in \cite{Pratt,Pratt2,Li}, the authors
found that the HBT radii hardly change if the peak of bulk viscosity
is small and the other parameters such as the equation of state,
initial condition and so on keep unchanged. However, in \cite{Torri}
the authors analyzed the stability of Navier-Stokes theory with the
bulk viscosity which has a large peak around $T_c$ and found that
there exists some inhomogeneous modes which will tear the system
into droplets. If this is the case, it will help us resolving the
interferometric data\cite{Torri2}. But there are some basic problems
with Navier-Stokes  theory as mentioned above and the authors did
not consider the effects of shear viscosity additionally. In this
paper we use the same method to study if this could happen in the
Israel-Stewart theory and what is the role of the shear viscosity.

\section{Israel-Stewart theory and linear perturbations}
The general hydrodynamic equations arise from the local conservation
of energy and momentum \cite{Kolb}
\begin{eqnarray}
\label{cons}
  \partial_\mu T^{\mu \nu}(x)&=&0,
\end{eqnarray}
where the energy-momentum tensor without heat conduction is
decomposed into the following form \cite{Chaud}
\begin{eqnarray}
\label{tmunu}
  T^{\mu\nu}&=& eu^{\mu}u^{\nu} - (p{+}\Pi)\Delta^{\mu \nu}
               + \pi^{\mu \nu}.
\end{eqnarray}
where $e$ and $p$ are the local energy density and thermal
equilibrium pressure, and $u^{\mu}$ is the 4-velocity of the energy
flow which obeys $u^\mu u_\mu = 1$. $\Pi$ is the bulk viscous
pressure, and $\Delta^{\mu\nu}=g^{\mu\nu}-u^{\mu}u^{\nu}$ is
transverse to the flow velocity, that is $\Delta^{\mu\nu}u_{\nu}=0$.
$\pi^{\mu \nu}$ is the traceless shear viscous pressure tensor. With
Eq.~(\ref{cons}) and Eq.~(\ref{tmunu}), we can get the evolution
equations of the energy density and the 4-velocity of energy flow.

In Israel-Stewart approach, the kinetic evolution equations of the
bulk pressure $\Pi$ and the traceless shear viscous tensor
$\pi^{\mu\nu}$ are \cite{Chaud}
\begin{eqnarray}
\label{Pi}
  D{\Pi}&=&-\frac{1}{\tau_{\Pi}}\big(\Pi+\zeta \nabla{\cdot}u\big),
\end{eqnarray}
\begin{eqnarray}
  D\pi^{\mu \nu} &=&-\frac{1}{\tau_{\pi}}\big(\pi^{\mu\nu}-2\eta\langle\nabla^\mu u^\nu\rangle\big)
\label{pi}
-\bigl(u^\mu\pi^{\nu\alpha} + u^\nu\pi^{\mu\alpha}\bigr)
      Du_\alpha,
\end{eqnarray}
where $D$ is the time derivative in the local fluid rest frame and
fulfills $D=u^{\mu}\partial_{\mu}$. The angular bracket notation is
defined by $\langle\nabla^\mu u^\nu\rangle=\frac{1}{2}(\nabla^\mu
u^\nu{+}\nabla^\nu u^\mu) -
\frac{1}{3}(\nabla{\cdot}u)\Delta^{\mu\nu}$. $\eta$ and $\zeta$
denote the bulk and shear viscous coefficient, respectively.
$\tau_{\pi}$ and $\tau_{\Pi}$ are the relaxation time for the bulk
pressure and the shear tensor, respectively. They can be related to
$\eta$ and $\zeta$ as follows \cite{Muronga}
\begin{eqnarray}
\label{rela}
\tau_{\pi}=2\eta\beta_2~~,~~~~~~~~~~~\tau_{\Pi}=\zeta\beta_0.
\end{eqnarray}
where $\beta_2$ and $\beta_0$ are the relaxation coefficients that
need to be calculated from other theories.

When hydrodynamics is applied to describe heavy ion collisions,
people always use the symmetry of the collision system to simplify
the evolution equations. Thus we have the scaling solution, the 1+1
solution and the 2+1 solution \cite{Muller,Roma1,Roma2,Song1}. Here
we focus on the 1+1 solution of the Israel-Stewart theory. For 1+1
solution, the system has boost-invariant longitudinal expansion and
transverse expansion in one dimension. For these geometries, it is
convenient to work in the co-moving and radial coordinates
$\tau,r,\phi,\eta$. We consider a small perturbation, and the
backgrounds $\epsilon_0(\tau), \pi^{\mu}_{\nu 0}(\tau), \Pi_0(\tau)$
evolve with time \bqa \label{total-e}
\epsilon(\tau,r)=\epsilon_0(\tau)+\delta\epsilon(\tau,r), \eqa 
\bqa \label{total-v} v(\tau,r)=\delta v(\tau,r), \eqa 
\bqa \label{total-pi} \pi^{\mu}_{\nu}(\tau,r)=\pi^{\mu}_{\nu
0}(\tau)+\delta\pi^{\mu}_{\nu}(\tau,r), \eqa 
\bqa \label{total-Pi} \Pi(\tau,r)=\Pi_0(\tau)+\delta\Pi(\tau,r).
\eqa

With Eqs.~(\ref{cons}$\sim$\ref{pi}) and
Eqs.~(\ref{total-e}$\sim$\ref{total-Pi}) the evolution equations of
backgrounds can be given as follows
\begin{eqnarray}
\label{back-e}
\partial_{\tau}\epsilon_0=-(\epsilon_0+p_0+\Pi_0)\frac{1}{\tau}
+\pi_{\eta 0}^{\eta}\frac{1}{\tau},
\end{eqnarray}
\begin{eqnarray}
\label{back-piee} \tau_{\pi 0}\partial_{\tau}\pi_{\eta
0}^{\eta}+\pi_{\eta 0}^{\eta} =\frac{4}{3}\frac{\eta_0}{\tau},
\end{eqnarray}
\begin{eqnarray}
\label{back-pirr} \tau_{\pi 0}\partial_{r}\pi_{r0}^{r}+\pi_{r0}^{r}
=-\frac{2}{3}\frac{\eta_0}{\tau},
\end{eqnarray}
\begin{eqnarray}
\label{back-Pi} \tau_{\Pi
0}\partial_{\tau}\Pi_0+\Pi_0=-\frac{\zeta_0}{\tau}.
\end{eqnarray}
We can also get the evolution equations of the perturbations
\begin{eqnarray}
\label{del-v} &&\!\!
[(\epsilon_0+p_0+\Pi_0-\pi_{r0}^r)\partial_{\tau}+\partial_{\tau}(p_0+\Pi_0)
-(\partial_{\tau}+\frac{1}{\tau})\pi_{r0}^r+\frac{1}{\tau}\pi_{\eta
0}^{\eta}]\delta v
\nonumber\\
&&~~~~~~~~~~~~~~~~~~~~~~~~~~+\partial_r(\delta
p+\delta\Pi)-(\partial_r
+\frac{2}{r})\delta\pi_{r}^{r}-\frac{1}{r}\delta\pi_{\eta}^{\eta}=0,
\end{eqnarray}
\begin{eqnarray}
\label{del-e} &&\!\!\!\!\!\!\!\!
[(\epsilon_0+p_0+\Pi_0)(\partial_r+\frac{1}{r})-\pi_{r0}^{r}(\partial_r-\frac{1}{r})
+\frac{1}{r}\pi_{\eta 0}^{\eta}]\delta
v+\partial_{\tau}\delta\epsilon
\nonumber\\
&&~~~~~~~~~~~~~~~~~~~~~~~~~~~~~~~~+\frac{1}{\tau}(\delta\epsilon+\delta
p+\delta\Pi)-\frac{1}{\tau}\delta\pi_{\eta}^{\eta}=0,
\end{eqnarray}
\begin{eqnarray}
\label{del-piee} \tau_{\pi
0}\partial_{\tau}\delta\pi_{\eta}^{\eta}+\delta\pi_{\eta}^{\eta}=
-\frac{\delta\tau_{\pi}}{\tau_{\pi
0}}(\frac{4}{3}\frac{\eta_0}{\tau}-\pi_{\eta 0}^{\eta})
-\frac{2}{3}\eta_0(\partial_r\delta v+\frac{\delta v}{r})
+\frac{4}{3}\frac{\delta\eta}{\tau},
\end{eqnarray}
\begin{eqnarray}
\label{del-pirr} \tau_{\pi
}\partial_{\tau}\delta\pi_r^r+\delta\pi_r^r=
\frac{\delta\tau_{\pi}}{\tau_{\pi
0}}(\frac{2}{3}\frac{\eta_0}{\tau}+\pi_{r0}^r)
-\frac{2}{3}\eta_0(-2\partial_r\delta v+\frac{\delta v}{r})
-\frac{2}{3}\frac{\delta\eta}{\tau},
\end{eqnarray}
\begin{eqnarray}
\label{del-Pi} \tau_{\Pi
0}\partial_{\tau}\delta\Pi+\delta\Pi=\frac{\delta\tau_{\Pi}}{\tau_{\Pi
0}} (\Pi_0+\frac{\zeta_0}{\tau})-\zeta_0(\partial_r\delta
v+\frac{\delta v}{r}) -\frac{\delta\zeta}{\tau}.
\end{eqnarray}
In order to get rid of the space-like derivatives, we can do the
Hankel transform (Fourier-Bessel transform) due to the geometry
system we use here\cite{Roma1}. After introducing \bqa
\label{delpirree}
\delta\tilde{\pi}=(\partial_r+\frac{2}{r})\delta\pi_r^r+\frac{1}{r}\delta\pi_{\eta}^{\eta}
\eqa we do the following Hankel transforms, \bqa \label{Ha1}
\delta{v}(\tau,r)=\int^\infty_0{\rm
d}kJ_1(kr)k\delta\tilde{v}(\tau,k), \eqa \bqa \label{Ha2}
\delta{\epsilon}(\tau,r)=\int^\infty_0{\rm
d}kJ_0k(kr)\delta\tilde{\epsilon}(\tau,k), \eqa \bqa \label{Ha3}
\delta{{\pi}}(\tau,r)=\int^\infty_0{\rm
d}kJ_1(kr)k\delta\tilde{{\pi}}(\tau,k), \eqa \bqa \label{Ha4}
\delta\pi^{\eta}_{\eta}(\tau,r)=\int^\infty_0{\rm
d}kJ_0(kr)k\delta\tilde{\pi}^{\eta}_{\eta}(\tau,k), \eqa \bqa
\label{Ha5} \delta{\Pi}(\tau,r)=\int^\infty_0{\rm
d}kJ_0(kr)k\delta\tilde{\Pi}(\tau,k), \eqa where $k$ is the wave
number. It stands for a homogeneous perturbation when $k$ equals
zero or an inhomogeneous perturbation when $k$ is not zero. Its
range for a realistic QGP fluid has been roughly estimated in
\cite{Koun}. With Eqs.~(\ref{del-v}$\sim$\ref{Ha5}), we can get the
evolution equations for $\delta\tilde{v}, \delta\tilde{\epsilon},
\delta\tilde{\pi_{\eta}^{\eta}}, \delta\tilde{\pi},
\delta\tilde{\Pi}$
\begin{eqnarray}
\label{Hdel-v}&&\!\!\!\!\!\!\!\!\!\!\!\!\!\!\!\!\!\!\!\!\!\!\!\!!\!\!\!\!\!
[(\epsilon_0+p_0+\Pi_0-\pi_{r0}^r)\partial_{\tau}+\partial_{\tau}(p_0+\Pi_0)
-(\partial_{\tau}+\frac{1}{\tau})\pi_{r0}^r+\frac{1}{\tau}\pi_{\eta
0}^{\eta}]\delta\tilde{v}
\nonumber\\
&&~~~~~~~~~~~~~~~~~~~~~~~~~~~~~~~~~~~~~~~~-kc_s^2\delta\tilde{\epsilon}
-k\delta\tilde{\Pi}-\delta\tilde{\pi}=0,
\end{eqnarray}
\begin{eqnarray}
\label{Hdel-e} (\epsilon_0+p_0+\Pi_0+\frac{1}{2}\pi^{\eta}_{\eta
0})\delta\tilde{v}+
(\partial_{\tau}+\frac{1+c_s^2}{\tau})\delta\tilde{\epsilon}
+\frac{1}{\tau}\tilde{\Pi}
-\frac{1}{\tau}\delta\tilde{\pi}^{\eta}_{\eta}=0,
\end{eqnarray}
\begin{eqnarray}
\label{Hdel-piee} &&\!\!\!\!\!\!\!\!\!\!\!\!\!\!
\frac{2}{3}\frac{\eta_0}{\tau_{\pi 0}}k\delta\tilde{v}
+\frac{1}{\tau_{\pi 0}^2}(\frac{4}{3}\frac{\eta_0}{\tau}-\pi_{\eta
0}^{\eta})
(\frac{\partial\tau_{\pi}}{\partial\epsilon})_0\delta\tilde{\epsilon}
-\frac{4}{3}\frac{1}{\tau\tau_{\pi
0}}(\frac{\partial\eta}{\partial\epsilon})_0\delta\tilde{\epsilon}
\nonumber\\
&&~~~~~~~~~~~~~~~~~~~~~~~~~~~~~~~~~~~~~~~~~~+(\partial_{\tau}+\frac{1}{\tau_{\pi
0}})\delta\tilde{\pi}_{\eta}^{\eta}=0,
\end{eqnarray}
\begin{eqnarray}
\label{Hdel-pirree} &&\!\!\!\!\!\!\!\!\!\!\!\!\!\!\!\!\!\!\!
\frac{4}{3}\frac{\eta_0}{\tau_{\pi 0}}k^2\delta\tilde{v}
-\frac{1}{2}\frac{1}{\tau_{\pi
0}^2}(\frac{4}{3}\frac{\eta_0}{\tau}-\pi_{\eta 0}^{\eta})
(\frac{\partial\tau_{\pi}}{\partial\epsilon})_0k\delta\tilde{\epsilon}
+\frac{2}{3}\frac{1}{\tau\tau_{\pi
0}}(\frac{\partial\eta}{\partial\epsilon})_0k\delta\tilde{\epsilon}
\nonumber\\
&&~~~~~~~~~~~~~~~~~~~~~~~~~~~~~~~~~~~~~~~~~~~~~~+(\partial_{\tau}
+\frac{1}{\tau_{\pi 0}})\delta\tilde{\pi}=0,
\end{eqnarray}
\begin{eqnarray}
\label{Hdel-Pi} \frac{\zeta_0}{\tau_{\Pi 0}}k\delta\tilde{v}
-\frac{1}{\tau_{\Pi 0}^2}(\frac{\zeta_0}{\tau}+\Pi_0)
(\frac{\partial\tau_{\Pi}}{\partial\epsilon})_0\delta\tilde{\epsilon}
+\frac{1}{\tau\tau_{\Pi
0}}(\frac{\partial\zeta}{\partial\epsilon})_0\delta\tilde{\epsilon}
+(\partial_{\tau}+\frac{1}{\tau_{\Pi 0}})\delta\tilde{\Pi}=0.
\end{eqnarray}

\section{Results}

In order to get numerical results, the values of viscosities and the
relaxation time are set up as described below. As to shear
viscosity, the strong coupling theory and the hydrodynamic and
transport model show that $\frac{\eta}{s}$ can not be too large
\cite{ADS,Roma2,Xuzhe}. Here we use two different values
$\frac{\eta}{s} = 0.02$ and 0.2 to see its effects. The relaxation
time of the shear viscosity is set to the Boltzmann gas result
$\tau_{\pi}=\frac{\eta}{s}\frac{6}{T}$ in our work. Although recent
results of SU(3) Yang-Mills theory show that the effects of a
potentially large bulk viscosity near $T_c$ are more subtle to
detect in spectral integrals\cite{Meyer} than previous work
suggested \cite{Meyer2}, but there are still some possibilities that
$\frac{\zeta}{s}$ becomes large around $T_c$\cite{PEAK}. So the
following parametrization as in \cite{Torri} is adopted to see the
effects of bulk viscosity on the stability 
\bqa \label{zos} \zeta=s(z_{pQCD}+\frac{z_0}{\sqrt{2\pi\sigma}}{\rm
exp}[-\frac{t^2} {2\sigma^2}]), \eqa 
where $t = T-T_c$, $\sigma = 0.01 T_c$, $z_{pQCD}\sim 10^{-3}$ and
different $z_0$ denotes different magnitudes of ${\zeta}/{s}$. For
the relaxation time of bulk viscosity, the parametrization similar
to the relaxation time of shear viscosity is employed \bqa
\label{rela-pa} \tau_{\Pi}=b\frac{\zeta}{\epsilon+p}, \eqa where
different $b$ is used to see the effect of relaxation time of bulk
viscosity.

To connect to relativistic heavy-ion collisions, the equation of
state in \cite{EOS} is adopted and the initial temperature $T_0$ is
set to be $0.34$ GeV. Firstly, let us see the effect of viscosity on
the evolution of background. Fig 1. shows the profiles of background
temperature for different $\eta/s$ and for different $z_0$ and $b$,
corresponding to different magnitudes and relaxation times of bulk
viscosity. Each plot consists of two different $\eta/s$. From the
left to the right, the plots represent $z_0$ = 0.0 and 0.1$~T_c$,
respectively. These plots show that the shear viscosity can slow
down the evolution of the source. We can also see that when $b$ is
large which means that the relaxation time of bulk viscosity is
large, bulk viscosity hardly have effects on the evolution of
temperature even when $\frac{\zeta}{s}$ has a large peak. But when
the relaxation time is small, for a larger magnitude of bulk
viscosity, we can see that the system stays at a nearly constant
temperature which is the same as the case found in Navier-Stokes
theory in \cite{Torri}. This effect is not hard to be understood.
Because the role of the relaxation time is to delay the appearance
of viscous forces \cite{Torri2}, bulk viscosity has effects on the
evolution of the source only after a time-scale $\tau_{\Pi}$. The
larger the relaxation time is, the longer time the bulk viscosity
needs to affect on the evolution of the system. Meanwhile, the
system is still evolving and it may be cooled down to be below $T_c$
where bulk viscosity can be negligible. So if the relaxation time is
large enough and the bulk viscosity has a small width, bulk
viscosity may have negligible effects on the evolution of
background. It is consistent with the results in \cite{Li}. In
\cite{Li} we found that the width of bulk viscosity has larger
effects on the evolution of the source than the magnitude of bulk
viscosity.

Fig 2. shows the profiles of the ratio of bulk pressure over the
equilibrium pressure of the background for different cases as in Fig
1. We can see that $\Pi/p$ has a peak when $\zeta/s$ has a peak
around $T_c$. And the fact that $\Pi/p$ exceeds one means that the
state is far away from equilibrium. It indicates that the matter is
not only hydrodynamically unstable, but also thermodynamically
unstable as stated in \cite{Muller} when the peak of bulk viscosity
is large and the relaxation time of bulk viscosity is small. We can
also see that shear viscosity will decrease $\Pi/p$ which indicates
that shear viscosity will weaken the instability that induced by
bulk viscosity. This behavior is also found in the next analysis
about linear perturbations.

Now we use Lyapunov direct method to study the stability of 1+1
solution of Israel-Stewart theory. The evolution equations of
perturbations Eqs.~(\ref{Hdel-v}$\sim$\ref{Hdel-Pi}) can be
rewritten in the following matrix formula 
\begin{eqnarray}
\partial_{\tau}{ \delta Y}={ A\delta Y},
\end{eqnarray}
where 
\bqa { \delta Y}=\left(
\begin{array}{cccc}
\delta \tilde{v}\\
\delta \tilde{\epsilon}\\
\delta \tilde{\pi}_{\eta}^{\eta}\\
\delta \tilde{\pi}\\
\delta \tilde{\Pi}
\end{array}\right).
\eqa 
When one uses the Lyapunov direct method, the Lyapunov function
should be given first. The Lyapunov function must be positive
definite. If it is a monotonically decreasing function, then the
solution is stable; if it is a monotonically increasing function,
then the solution is unstable. The more detailed description about
this method can be found in\cite{Koun,Deni2}. Here we assume the
Lyapunov function is $V=\delta Y^T\delta Y$. Then the evolution
equation of the Lyapunov function $V$ is \bqa
\partial_{\tau}V=\delta Y^T(A^T+A)\delta Y,
\eqa after a short derivation, we can get \bqa
\lambda_{min}V\leq\partial_{\tau}V\leq\lambda_{max}V \eqa where
$\lambda_{\rm max}$ and $\lambda_{min}$ are the largest and smallest
eigenvalues of ${ A}$ + ${A^T}$, respectively. It can be shown that
the solution is stable when $\lambda_{\rm max}\leq 0$ and is
unstable when $\lambda_{\rm min}\ge 0$. Fig. 3 and Fig. 4 show the
values of $\lambda_{\rm max}$ and $\lambda_{\rm min}$ for $k=0$ and
$k=3$, respectively. Each figure has differen cases as stated for
Fig. 1. We can see that neither the stable region nor the unstable
region can be determined because $\lambda_{\rm max}$ is always
larger than zero and $\lambda_{\rm min}$ is always smaller than
zero. As for the scaling solution, the unstable regions also can not
be found but the stable regions can be determined \cite{Deni2}.
These plots show that $\lambda_{\rm max}$ and $\lambda_{\rm min}$
have rapid change when the bulk viscosity is large and the
relaxation time is small. We can also see that shear viscosity will
delay this kind of rapid change and reduce the magnitude of
$\lambda_{\rm max}$ and $\lambda_{\rm min}$. This phenomenon is the
same for different $k$, but the magnitude of $\lambda_{\rm max}$ and
$\lambda_{\rm min}$ is larger with larger $k$.

We can see from Fig. 3 and Fig. 4 that the absolute value of
$\lambda_{\rm max}$ and $\lambda_{\rm min}$ increase rapidly which
means the growing and damping rates of perturbations increase
rapidly due to the large peak and small relaxation time of bulk
viscosity. Therefore the perturbations may rapidly grow to a value
comparable with the background. So they will break local homogeneity
and play an important role in the subsequent evolution of the
system. The created inhomogeneities have no reinteractions. It is
possible that isolated fragments will be created and move away from
each other. In \cite{Torri}, the authors argued that this may be a
reason that the source will be clusterized and then decoupled as the
fireballs. By adding a further free parameter which is the cluster
size to the system, this mechanism may solve the HBT
problem\cite{Torri2}. In the Israel-Stewart theory this phenomenon
may also happen when the peak of $\zeta/s$ is large, $\tau_\Pi$ is
small and $\zeta/s$ is not too large.

\section{Conclusions}
In summary, the stability problem of the 1+1 solution of
Israel-Stewart theory is studied.  Firstly, the evolution of the
temperature and the ratio of the bulk pressure over the equilibrium
pressure are studied. We find that both shear and bulk viscosity
slow down the evolution of temperature. And the ratio of bulk
pressure over the equilibrium pressure ($\Pi/p$) will exceed one
with a large peak of bulk viscosity, which indicates the state is
unstable. The shear viscosity reduces the magnitude of $\Pi/p$ to
weaken the effects of bulk viscosity. Then using Lyapunov direct
method, we can not determine the stable or unstable regions. We also
find the phenomenon which may drive the source to clusterize which
is similar to that in Navier-Stokes theory. However, this phenomenon
will happen only when the peak of bulk viscosity is large enough,
the relaxation time of the bulk viscosity is small and the shear
viscosity is not too large.

\vspace{0.5cm} Acknowledgments:
 This work is supported by the
National Natural Science Foundation of China (Grant Nos 10775167 and
10705044), the Knowledge Innovation Project of Chinese Academy of
Sciences (Grant Nos KJCX2-YW-A14), the Startup Foundation for the
CAS Presidential Scholarship Award of China (Grant No 29010702), the
Shanghai Development Foundation for Science and Technology under
Grant Nos. 09JC1416800.

\bibliographystyle{model1a-num-names}
\bibliography{<your-bib-database>}



\newpage

Figure 1: Evolutions of background temperature for different shear
and bulk viscosity sets. See texts for details.

Figure 2: $\Pi/p$ as a function of $T/T_c$ for different shear and
bulk viscosity sets. See texts for details.

Figure 3: $\lambda_{max}$ and $\lambda_{min}$ for $k = 0$ with
different shear and bulk viscosity sets. See texts for details.

Figure 4: $\lambda_{max}$ and $\lambda_{min}$ for $k = 3$ with
different shear and bulk viscosity sets. See texts for details.

\newpage

\begin{figure*}[pt]
\epsfig{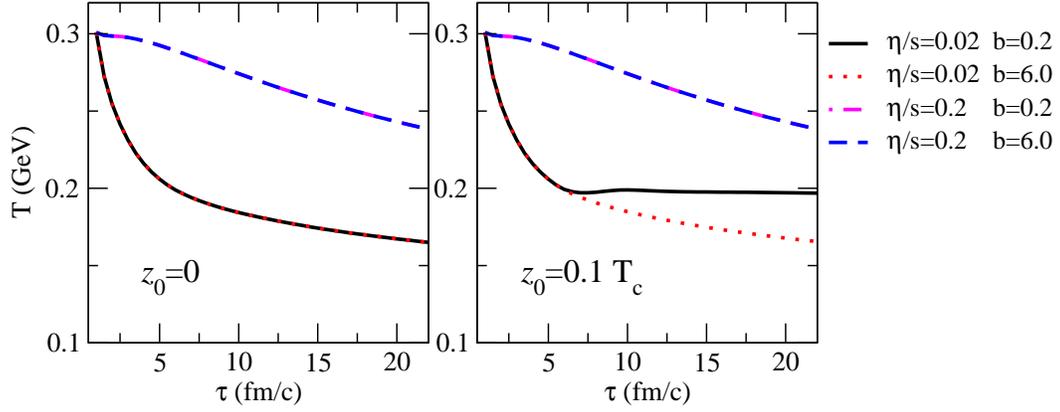} \caption{\label{tplot}
Evolutions of background temperature for different shear and bulk
viscosity sets. See texts for details.}
\end{figure*}

\begin{figure*}[t]
\epsfig{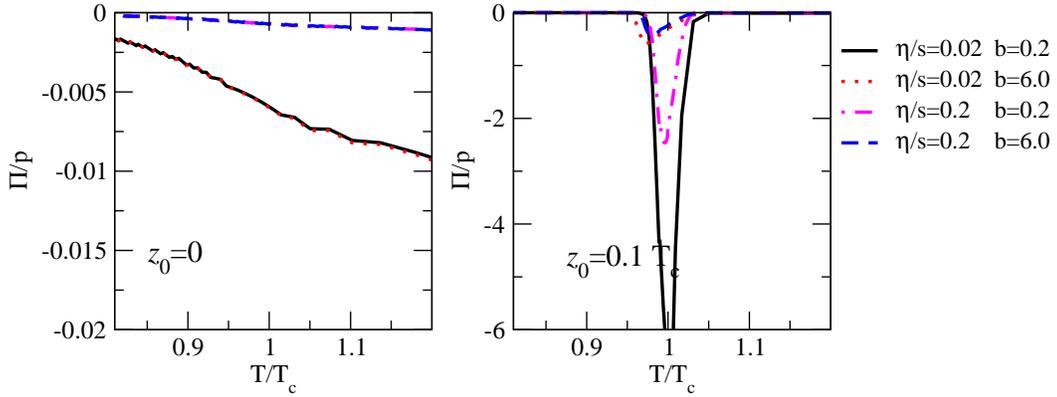}
\caption{\label{lamplot} $\Pi/p$ as a function of $T/T_c$ for
different shear and bulk viscosity sets. See texts for details. }
\end{figure*}

\begin{figure*}[t]
\epsfig{width=14cm,clip=1,figure=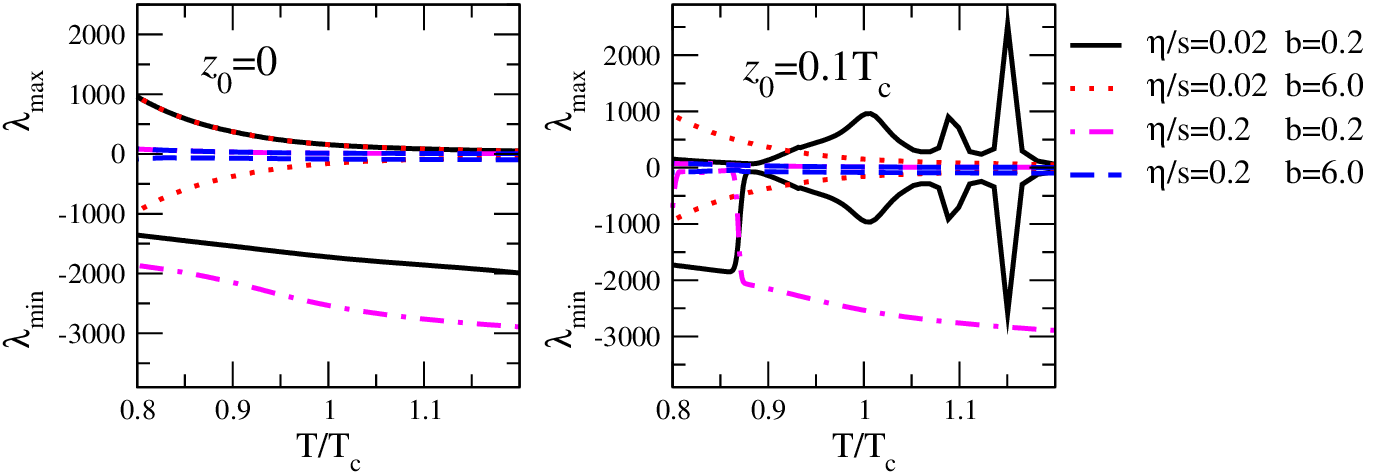}
\caption{\label{ratplot} $\lambda_{\rm max}$ and $\lambda_{\rm
min}$ for $k=0$ with different shear and bulk viscosity sets. See
texts for details.}
\end{figure*}
\begin{figure*}[t]
\epsfig{width=14cm,clip=1,figure=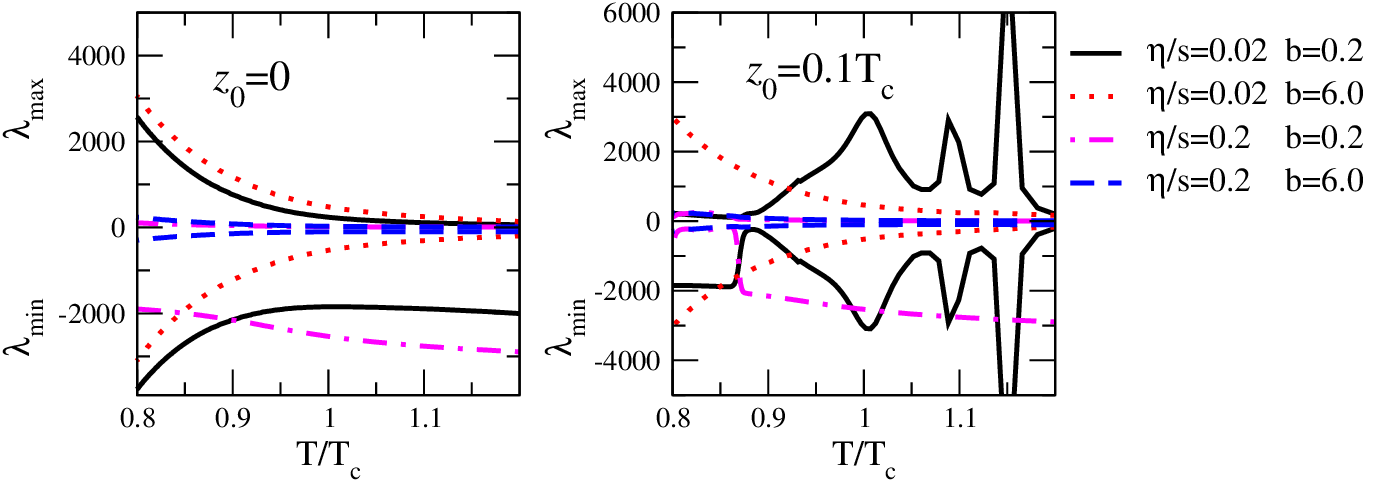}
\caption{\label{ratplot} $\lambda_{\rm max}$ and $\lambda_{\rm
min}$ for $k=3$ with different shear and bulk viscosity sets. See
texts for details.}
\end{figure*}
\end{document}